\documentclass[aps,pre,twocolumn,amsmath,amssymb,showpacs,amsfonts]{revtex4}
\usepackage{epsfig}
\usepackage{graphicx}
\usepackage{dcolumn}
\usepackage{bm}

\begin{document}

\title{Effect of distinguishability of patterns of collisions of particles in a non-equilibrium chaotic system}

\author{Alexander Jonathan Vidgop$^1$}
\author{Itzhak Fouxon$^{2}$}

\affiliation{$^1$ Am haZikaron Institute, Tel Aviv 64951, Israel}
\affiliation{$^2$ Raymond and Beverly Sackler School of Physics and Astronomy,
Tel-Aviv University, Tel-Aviv 69978, Israel}
\begin{abstract}

We follow the time sequence of binary elastic collisions 
in a small collection of hard-core particles. 
Intervals between the collisions are characterized by the numbers of collisions of different pairs in a given time. It was shown previously that due to the ergodicity these numbers grow with time as a biased random walk. We show that this implies that
for a typical trajectory in the phase space each particle has "preferences" that are stable during indefinitely long periods of time. During these periods the particle collides more with certain particles and less with others. Thus there is a clearly distinguishable pattern of collisions of the particle with other particles, as determined by its initial position and velocity. The effect holds also for the dilute gas with arbitrary short-range interactions allowing for experimental testing. It is the mechanical counterpart to the classical probabilistic observation that "in a population of normal coins the majority is necessarily maladjusted" \cite{Feller}.

 
\end{abstract}
 \maketitle

Chaos makes individual trajectories of closed mechanical systems generally inaccessible to the direct study. Correspondingly, the ergodic theory considers instead of the evolution of one initial point in the phase space, the evolution of finite volumes in the allowed region of the phase space (the microcanonical ensemble). The theory typically studies the given mechanical system in terms of integrals over initial conditions. In particular, these integrals define the different-time correlations of functions on the phase space (mechanical variables) and the central limit theorem (CLT), which is based on the decay of these correlations in time \cite{SinaiBook}. The latter says roughly that for the ergodic systems the mechanical variables become independent at significantly separated instants of time, where the statistical independence is defined by the microcanonical ensemble. Then the quantities representable as time integrals of a mechanical variable must perform a random walk at large times. It follows that a typical individual trajectory of the system must be such that on the coarse-grained time-scale the time-averages of all its functions are statistically indistinguishable from the random walk. Then by a choice of functions one may reveal a wealth of physically relevant information on the system.
The most immediate example is the trajectory itself. That can be represented as a time integral of the flow in the phase space. Then, assuming the decay of the flow's correlations, at large times the trajectory performs diffusion over the allowed region of the phase space, the fact that underlies the whole approach of the ergodic theory to the equilibrium statistical mechanics.

In this Letter we use a different function of the trajectory to get insight into the details of collisions in the dilute gas with arbitrary short-range interactions. The main question we address is: how much the particles collide with each other within a finite interval of time $t$? To make the question meaningful we consider systems with a moderate number of particles where each particle collides many times with every other particle in the system at a reasonable $t$. We inquire if these collisions are uniform so that every particle collides with every other particle more or less the same number of times within $t$. We find that typically the numbers of the particle's collisions with other particles deviate from uniformity strongly. Typical initial conditions for the position and the velocity of the particle define the evolution where the particle collides with other particles non-equal numbers of times with the difference typically growing in time. Thus chaos implies a rather counterintuitive property of the system which meaning can be made more detailed for hard-core interactions. In this case the particles bounce off each other instantaneously. Chaos can be characterized by noting that a slight change in the collision against the convex surface of the other particle may result in the large change of the particle's position in the future. Then our result says that sequences of collisions where the particles collide with each other more or less the same number of times are highly improbable (the initial conditions that correspond to such trajectories occupy a small fraction of the allowed phase space volume). Rather, generically, the particle collides more with certain particles and less with others during very long intervals of times. These "preferences" sooner or later are changed to other preferences so that eventually all possible preferences will appear, however the expectation time to these events is infinite. 

The study assumes that the evolution consists of the free motion of the particles interrupted by the binary collisions.
For hard-core interactions where the collisions are instantaneous, this frame holds also in the dense fluid phase, to which the results can be extended. We designate the number of collisions of the pair of particles $i$ and $j$ within time $t$ by $N_{ij}(t)$. These 
describe times between the collisions and on average $N_{ij}(t)$ are equal to the ratio of $t$ to the mean free time between the collisions of the pair. These numbers were recently represented as the time integrals of the mechanical variable \cite{JI}. Using facts that hold for the random walk with probability one, predictions were made for individual trajectories.

We use hard balls (disks) as a basic model of study. Many statements of the ergodic theory, that are often assumed for the systems studied in statistical mechanics, are actually proved for this model \cite{SinaiBook,Dorfman}. The famous Boltzmann-Sinai hypothesis states that systems of an arbitrary number $N\geq 2$ of elastic hard balls in a $d-$dimensional box with periodic boundary conditions (torus), $d\geq 2$, are ergodic in the phase space region where the trivial conserved quantities of the system are constant \cite{Sinai}. Today this hypothesis can be considered as "almost proved", see e. g. \cite{Simanyi} and references therein, and also \cite{Szasz,Mulero}. The ergodicity does not demand a large $N$: already a system of two disks in a $2-$dimensional torus is ergodic \cite{Sinai2}. Thus the systems with a moderate number of particles considered here can be assumed ergodic.

The motion of hard balls generates an ordered sequence of pairs, say $(1, 2)$ $(5, 6)$ $(7, 8)...$, meaning that first the particles $1$ and $2$ collided, then $5$ and $6$, then $7$ and $8$ and so on. In particular, this sequence is the basic result of the event-driven simulations (EDS) of this system \cite{AT}.
As the list is determined by the Newton law, the subsequent pairs in the sequence are not independent. Nevertheless, due to chaos, one expects that well-separated pairs in the sequence are independent, so on a large time scale the above sequence gets close to the sequence of random pairs where the next pair is picked at random, cf. \cite{Ornstein,GalOrn}. This expectation was formalized in \cite{JI} where it was shown that $N_{ij}(t)$ are statistically the same as the biased random walk on the coarse-grained time-scale. Below we generalize this result to the dilute gas with arbitrary short-range interactions. A way to define the sequence of non-instantaneous collisions, that may overlap in time in this case, is described below. 
To establish the properties of the collisions described above, we use
the facts of the probability theory \cite{Feller} that give robust expression to the fact that the behavior of a single realization of the random walk is very different from the ensemble average. The main observation is that for arbitrarily large number $n$ of the steps of the random walk, the most probable fraction of time that the random walk did not change its sign is one. This is in sharp contrast to the intuitive expectation that in the limit of large $n$ the fraction of time that a long walk is positive (negative) is about $1/2$. In fact, the fraction $1/2$ is the least probable, which produces many surprising phenomena \cite{Feller}. The application to the gas seems to be missing from the literature.

The analysis is performed either for dilute gases of particles or for the fluid of hard balls, that collide in the three-dimensional box with periodic boundary conditions (the generalization to hard walls is straightforward). For the gas we neglect triple
collisions using the diluteness. The basic observation is that $N_{ij}(t)$ can be represented as a time-integral of a certain function $F(\bm r_{ij}, \bm v_{ij})$ on the phase space,
\begin{eqnarray}&&\!\!\!\!\!\!\!
N_{ij}(t)=\int_0^t \xi_{ij}(t') dt',\ \ \xi_{ij}(t)=F\left[\bm r_{ij}(t), \bm v_{ij}(t)\right], \label{definition}
\end{eqnarray}
where $\xi_{ij}(t)$ is the collision rate of the particles $i$ and $j$. Above $\bm r_{ij}$ and $\bm v_{ij}$ are the relative coordinate and velocity of the particles $i$ and $j$, respectively. We consider $F(\bm r, \bm v)$. We assume that $k-$th binary collision of particles $i$ and $j$ starts at the time $t_{ij}^k$, when the two particles approach each other to a fixed threshold distance $d$ and ends at time $t_{ij}^k+\tau_{col}$, when they cross the threshold while separating ($\tau_{col}$ depends on the collision's details). For hard balls $d$ is the particles' diameter and $\tau_{col}=0$. Thus we assume that the interactions have short-range, while for the long-range interactions the analysis needs refinement, cf. \cite{Landau}. Using any function $\chi(r)$ that is constant for $r\leq d$, where it equals $1/2$, and
vanishes in a smooth way for $r$ slightly larger than $d$, we represent $F(\bm r, \bm v)=-|\bm r\cdot\bm v|\chi'(r)/r$, cf. \cite{JI}. We have
\begin{eqnarray}&&\!\!\!\!\!\!\!
-\int |\bm r(t)\cdot\bm v(t)|\chi'\left[r(t)\right] dt/r(t)=-\int \chi'\left[r(t)\right] |\dot{r}(t)|dt.\nonumber
\end{eqnarray}
Performing the integral from times slightly smaller than $t_{ij}^k$, where $\chi\left[r_{ij}(t)\right]=0$ and $\dot{r}_{ij}<0$, to times slightly larger than $t_{ij}^k+\tau_{col}$, where $\chi\left[r_{ij}(t)\right]$ again equals zero with $\dot{r}_{ij}>0$, we find that the contribution of these times to the above integral is unity. Here we assumed that the collision occurs so that $r_{ij}(t)$ monotonically decreases to the minimum and then starts to increase monotonically. This assumption is valid for central forces, while appropriate generalizations can be considered in other cases. Thus
$F(\bm r, \bm v)=-|\bm r\cdot\bm v|\chi'(r)/r$ can be used in
Eq.~(\ref{definition}).
For our purposes only the existence of $F(\bm r, \bm v)$ is needed, that allows to use the machinery of the ergodic theory to study the collision rate $\xi_{ij}(t)$. In doing so we will assume that the non-analyticity in $F(\bm r, \bm v)$ due to the presence of $|\bm r\cdot\bm v|$ factor does not prevent one from using the results of the ergodic theory. This assumption is used in the kinetic theory to find the collision frequency \cite{Landau} and it seems sound. We expect a proof can be derived for the hard balls, which is left for future work.

The first conclusion from the ergodic theory is the (expected) existence of the average collision rate
\begin{eqnarray}&&\!\!\!\!\!\!\!\!\!\!
\nu=\lim_{t\to\infty}N_{ij}(t)/t=\lim_{t\to\infty}\int_0^t \xi_{ij}(t') dt'/t=\langle F(\bm r, \bm v)\rangle,
\end{eqnarray}
where we omitted $i$, $j$ in the last term since $\nu$ is the same for all pairs of particles and it is equal to the inverse average
time between their collisions.
The angular brackets designate the equilibrium averages with respect to the microcanonical ensemble defined by the trivial conserved
quantities of the system (for torus the energy and the momentum). Thus the statistics is defined by picking the initial condition at random in the
allowed region of the phase space. While the above applies to a single trajectory, below we address the trajectories statistically.

It is the deviations of $\xi_{ij}(t)$ from its mean $\nu$ determine the effect discussed here. The fluctuations produce 
\begin{eqnarray}&&\!\!\!\!\!\!\!
\lim_{t\to\infty}\frac{\left\langle\left[N_{ij}(t)-\nu t\right]\left[N_{mn}(t)-\nu t\right]\right\rangle}{t}=2\Gamma_{ij, mn},\ \ \ \Gamma_{ij, mn}
\nonumber\\&&\!\!\!\!\!\!\!
\!=\!\lim_{t\to\infty}\Gamma_{ij, mn}(t),\ \ \Gamma_{ij, mn}(t)\!\equiv\!
\int_0^t\left[\langle \xi_{ij}(0)\xi_{mn}(t)\rangle\!-\!\nu^2 \right]dt.\nonumber
\end{eqnarray}
One can write $\Gamma_{ij, mn}(t)=\langle \xi_{ij}(0)N_{mn}(t)\rangle-\langle \xi_{ij}(0)\rangle \langle  N_{mn}(t)\rangle$ so $\Gamma_{ij, mn}(t)$ measures the deviation of $N_{mn}(t)$ from the mean $\nu t$ for evolutions that start from the initial conditions where the particles $i$ and $j$ collide. We assume that $\Gamma_{ij, mn}$ exist, so
$\xi_{ij}(t)$ effectively has a finite correlation time $\tau_{cor}<\infty$, cf. \cite{Szasz,JI}. This assumption is justified in the considered
three-dimensional case, since one expects the correlations of $\xi_{ij}$
to decay at large times as $t^{-D/2}$, where $D$ is the space dimension \cite{EW,PR}. It follows from the CLT that at $t\gg\tau_{cor}$ the distribution of $N_{ij}(t)$ is Gaussian and determined completely by $\nu$ and $\Gamma_{ij, mn}$.
This allows to substitute $\xi_{ij}(t)$ by $\nu+\zeta_{ij}$, where $\zeta_{ij}(t)$ is a white-noise,
\begin{eqnarray}&&\!\!\!\!\!\!\!
\dot{N}_{ij}=\nu+\zeta_{ij},\ \ \langle\zeta_{ij}(t)\zeta_{mn}(t') \rangle=2\Gamma_{ij, mn}\delta(t-t'). \label{Langevin}
\end{eqnarray}
We note that the use of the CLT above neglects the tails in the distribution. Though these tails can be even algebraic, see e. g. \cite{Sanders}, this is not a limitation for the analysis below, that concerns the most probable events.

The effective Langevin equations (\ref{Langevin}) were introduced in \cite{JI} for the system of hard balls. They produce statistically the same  process as in Eq.~(\ref{definition}), on a coarse-grained time-scale much larger than $\tau_{cor}$. For hard balls the result applies to the fluid-like state of the system, where all the particles are well-mixed and all pairs collide (the situation where some particles are trapped would produce infinite $\tau_{cor}$). As the original process $N_{ij}(t)$ is discrete and it grows in jumps, a more intuitive form of the result can be obtained by embedding Eq.~(\ref{Langevin}) in a biased random walk with parameters defined by $\nu$ and $\Gamma_{ij, mn}$.

\begin{figure}
\includegraphics[width=7.9 cm,clip=]{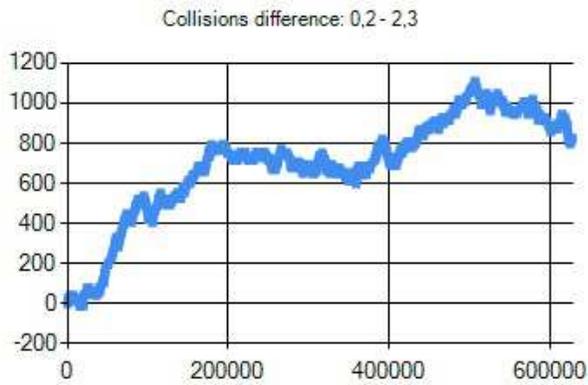}
\caption{Shown is the plot of the difference of the number of collisions of particle $2$ with particle $0$ and particle $3$. The
simulation involved $3.5\times 10^6$ events, consisting either of particles' collisions between themselves or of passing a wall. Five particles were introduced into the box of size $100\times 100$ with random initial conditions, where in the used units the r. m. s. velocity was set to $100$. The particles' radius equals $2$. Periodic boundary conditions were used. Each particle made about $7\times 10^4$ collisions with other particles.}
\label{Figure}
\end{figure}

The Langevin equations (\ref{Langevin}) produce non-trivial physical conclusions \cite{JI}. The
effective correlations and anti-correlations between the particles are described by the 
differences ${\tilde N}_{ij, mn}(t)\equiv N_{ij}(t)-N_{mn}(t)$ that obey the equations of the usual Brownian motion
\begin{eqnarray}&&\!\!\!\!\!\!\!
\dot{{\tilde N}}_{ij, mn}=w_{ij, mn},\ \ w_{ij, mn}=\zeta_{ij}-\zeta_{mn},
\label{diff}\\&&\!\!\!\!\!\!\!
\langle w_{ij, mn}(t)w_{ij, mn}(t')\rangle=4\left[\Gamma_{ij, ij}-\Gamma_{ij, mn}\right]
\delta(t-t'), \label{Brownian}
\end{eqnarray}
The statements that hold for the Brownian motion with probability one were applied to derive the deterministic predictions about
the behavior of the system \cite{JI}, concerning the repetitions of the rare events ${\tilde N}_{ij, mn}(t)=0$ that have infinite expectation time.
Here we consider the typical behavior of ${\tilde N}_{ij, mn}(t)$ providing the bulk of the particles' correlations. 
For a typical evolution of ${\tilde N}_{ij, in}(t)$ governed by Eq.~(\ref{diff}), possibly after relatively small initial fluctuations, ${\tilde N}_{ij, in}(t)$ will perform a long and large excursion to, say, positive values (if the excursion is to negative values, consider instead 
${\tilde N}_{in, ij}[t]=-{\tilde N}_{ij, in}[t]$). This excursion means that the particle $i$ will start colliding more with particle $j$ than with particle $n$. As the excursion progresses the difference between the numbers of collisions grows. A typical plot, which details will be discussed later, is shown in Fig. \ref{Figure}. It shows that the difference of the numbers of collisions of the particle $2$ with the particle $0$ and particle $3$ grows in time steadily with small fluctuations. The observer will see that during a relatively long period of time (during which each pair collided about $7\times 10^4$ times), the particle $2$ has a steady inclination to collide more with particle $0$ than with particle $3$.
This looks like there is an effective attraction between the particles $2$ and $0$.
This attraction is not due to the force but due to different initial conditions that generally produce patterns with widely different $N_{ij}(t)$. 
This is reinforced by the fact that if one waits long enough the random walk ${\tilde N}_{20, 23}(t)$ will return to the origin ${\tilde N}_{20, 23}(t_*)=0$ at $t=t_*$ and the effective attraction changes to either attraction or repulsion with equal probability. This is a non-equilibrium effect pertaining to finite times of observation, rather than the infinite observation time that corresponds to equilibrium in the frame of the ergodic theory. However, the waiting time $t_*$ can be arbitrarily large and its probability density function (PDF) has a power-law tail $t_*^{-3/2}$  see e. g. \cite{Feller}. The average is infinite, $\langle t_*\rangle=\infty$, and the observer is likely to see
stable differences in the patterns of particles' collisions during the whole time of the experiment. 
Generally one has that the particle $i$ has "preferences" and collides more with particle $j$, than with particle $n$. The preferences hold for long times and can be considered as a long-living, quasi-stationary excitation of the system.

The quantitative description in terms of relative fractions of the phase space volume of the initial conditions that produce given preferences is convenient to formulate by embedding ${\tilde N}_{ij, mn}(t)$ in the discrete random walk and
using the laws closely related to the arcsine law \cite{Feller}. Assuming ${\tilde N}_{ij, mn}(t)$ changes by unit steps fixes uniquely
the time-interval $\Delta t$ between the jumps from the diffusion coefficient of $w_{ij, mn}(t)$ in Eq.~(\ref{Brownian}). Then for the sufficiently
large number $2n$ of the steps of the random walk, the arcsine law implies that for $0<x<1$ the probability that $xn$ time units
the walk was positive, and $(1-x)n$ was negative, tends to an $n-$independent limit $(2/\pi)\arcsin\sqrt{x}$.
The law implies that ${\tilde N}_{ij, mn}(t)$ keeps its sign in the interval
$(t/2, t)$ with probability $1/2$, independently of how large $t$ is. Furthermore, the most probable fractions of time during which a random walk is positive
are either close to zero or to one, and not to $1/2$. The probability that ${\tilde N}_{ij, mn}(t)$ changed its sign $k$
times decreases with $k$ and it is most probable that ${\tilde N}_{ij, mn}(t)$ never changed its sign for arbitrarily large $t$.  A
closely associated result is the statement that the probability that the random walk changes sign less than $x\sqrt{n}$ times
tends to a finite limit at $n\to\infty$. Thus the frequency of the changes decreases with $n$ and the duration of
intervals of constant sign increases in length. Moreover, very early and very late sign changes are the most probable ones.
These features are closely related to the $t_*^{-3/2}$ tail: the return may take extremely long
times while the random walk performs the corresponding large excursion \cite{Feller}.

The same laws apply to the difference of the total numbers $N_i(t)$ of collisions of different particles, as it is clear from the definition $N_i(t)=\sum_{j\neq i}N_{ij}(t)$. For a typical trajectory certain particles collide more, while others collide less for long periods of time. To describe different preferences of the particles one may use the $K(K-1)/2$ dimensional random walk $N_{ij}(t)-\nu t$, where $K$ is the number of particles in the system. Finally, to show the same behavior holds in $D=2$ where $\tau_{cor}=\infty$ we performed the EDS, see Fig.~\ref{Figure}. Though the difference ${\tilde N}_{ij,mn}$ is small relatively to $N_{ij}$ itself, it is systematic, which makes it physically relevant.
We also verified that similar behavior holds for the hard walls' boundary conditions.

The ergodic theory was used to obtain a rather non-trivial information on the behavior of the individual trajectory of the dilute gas or the fluid of hard-core particles. We showed that the characteristic feature of the chaos - the decay of the correlations in time - produces a special pattern in the time sequence of the particles' collisions. A single realization of the system's evolution has long periods of time during which all particles have their own preferences, colliding more with certain particles and less with others. Thus the initial positions and velocities of the particles determine for each one of them a unique distinguishable pattern of collisions with other particles. This behavior of the mechanical trajectories of the system does not look easily derivable by other means.

The effect applies to a system with any $K$, as long as it is ergodic. Its practical observation (numerical or experimental) however demands a moderate number of particles in view of the large numbers of collisions of the same pairs of particles involved \cite{JI}. For large $K$, the system can be divided into cells, such that the collisions occur mostly within the cell \cite{AT} and the results could apply locally in space, as it is to be studied.

The described effect of the preferences is an example of the importance of distinguishing the behavior of the single trajectory and ensembled-averaged behavior for non-equilibrium systems, cf. \cite{Barkai}. An important question it poses is to obtain the direct explanation from the mechanical point of view. Qualitatively the same physics underlies the effect and the chaotic divergence of the trajectories. Collisions off the convex surfaces of other particles are unstable with respect to small changes making the trajectories with uniform $N_{ij}$ too special. The sensitivity to small changes in the initial conditions implies that the overall number of collisions' patterns grows fast with time, making the preferences - large deviations of collisions' patterns from the mean - typical. Similar effects are expected in billiards and the Lorentz gas. 

It is feasible to test our predictions for the dilute gas with arbitrary short-range interactions experimentally. This would give a test of ergodicity for small systems.

We thank N. Simanyi, N. Chernov, J. R. Dorfman, S. Fishman and T. Gotoh for discussions.



\end{document}